\documentstyle[11pt]{article}
\begin{document}
\thispagestyle{empty}
\rightline{UOSTP-00-106}
\rightline{{\tt hep-th/0008204}}

\

\def\tr{{\rm tr}\,} \newcommand{\beq}{\begin{equation}}
\newcommand{\eeq}{\end{equation}} \newcommand{\beqn}{\begin{eqnarray}}
\newcommand{\eeqn}{\end{eqnarray}} \newcommand{\bde}{{\bf e}}
\newcommand{\balpha}{{\mbox{\boldmath $\alpha$}}}
\newcommand{\bsalpha}{{\mbox{\boldmath $\scriptstyle\alpha$}}}
\newcommand{\betabf}{{\mbox{\boldmath $\beta$}}}
\newcommand{\bgamma}{{\mbox{\boldmath $\gamma$}}}
\newcommand{\bbeta}{{\mbox{\boldmath $\scriptstyle\beta$}}}
\newcommand{\lambdabf}{{\mbox{\boldmath $\lambda$}}}
\newcommand{\bphi}{{\mbox{\boldmath $\phi$}}}
\newcommand{\bslambda}{{\mbox{\boldmath $\scriptstyle\lambda$}}}
\newcommand{\ggg}{{\boldmath \gamma}} \newcommand{\ddd}{{\boldmath
\delta}} \newcommand{\mmm}{{\boldmath \mu}}
\newcommand{\nnn}{{\boldmath \nu}}
\newcommand{\diag}{{\rm diag}}
\newcommand{\bra}[1]{\langle {#1}|}
\newcommand{\ket}[1]{|{#1}\rangle}
\newcommand{\sn}{{\rm sn}}
\newcommand{\cn}{{\rm cn}}
\newcommand{\dn}{{\rm dn}}
\newcommand{\tA}{{\tilde{A}}}
\newcommand{\tphi}{{\tilde\phi}}
\newcommand{\bpartial}{{\bar\partial}}
\newcommand{\br}{{{\bf r}}}
\newcommand{\bx}{{{\bf x}}}
\newcommand{\bk}{{{\bf k}}}
\newcommand{\bq}{{{\bf q}}}
\newcommand{\bQ}{{{\bf Q}}}
\newcommand{\bp}{{{\bf p}}}
\newcommand{\bP}{{{\bf P}}}
\newcommand{\thet}{{{\theta}}}
\renewcommand{\thefootnote}{\fnsymbol{footnote}}
\

\vskip 0cm
\centerline{\Large
\bf 
 Exact Solutions of Multi-Vortices and False Vacuum Bubbles }
\centerline{\Large 
\bf 
in
Noncommutative Abelian-Higgs Theories}
\vskip .2cm

\vskip 1.2cm
\centerline{ Dongsu Bak\footnote{Electronic Mail: dsbak@mach.uos.ac.kr}
}
\vskip 8mm
\centerline{Physics Department, 
University of Seoul, Seoul 130-743 Korea}
\vskip0.3cm


\vskip 1.2cm
\begin{quote}
{
We consider the noncommutative Abelian-Higgs theory  
and construct new type of exact multi-vortex 
solutions that solve
the static equations of motion. 
They in general do not follow from the BPS equations;
only for some specific values of parameters, they satisfy
the BPS equations saturating the Bogomol'nyi bound.
We further consider the Abelian-Higgs 
theory with more complicated scalar potential allowing unstable 
minima and construct  exact solutions of
noncommutative false vacuum bubble
with integer 
magnetic flux. The classical stability of 
the solutions is discussed.
}
\end{quote}


\newpage


The ordinary  Abelian-Higgs theory 
possesses all the essential 
ingredients leading to the spontaneous symmetry breaking 
via the condensation of the scalar field, serving as
a model for superconductivity.
The Maxwell field as well as the Higgs field
 become massive 
and magnetic fluxes are confined. 
The theory allows the 
topological vortex solution carrying $m$ unit magnetic 
fluxes\cite{nielsen}.
Especially in the critical case, there exist  static 
multi-vortex solutions saturating the BPS bound. However, 
until now, no closed form vortex solutions are found in 
this theory. Recently, some aspects of the 
Nielsen-Olesen vortices
in the noncommutative Abelian-Higgs theory
have been investigated via the BPS equations\cite{jatcar,waida}
and  related works have appeared\cite{polychronakos,gross}.

In this letter, we shall consider the noncommutative Abelian-Higgs model. 
(The gauge symmetry is in fact non-Abelian due to the noncommutative 
nature of the theory.) Our key finding is new types of 
exact multi-vortex 
solutions. When  the mass  parameter of the Higgs field takes a 
certain value, these solutions saturate the BPS bound and the 
classical stability is guaranteed in this case. We discuss
the detailed characteristics of the vortex solutions mainly focused on
the static properties. In particular, the gauge fields arising 
here give  localized integer-valued fluxes, which are also
static solutions of the noncommutative Maxwell 
theory\cite{polychronakos}.

We further consider the Abelian-Higgs 
theory with more complicated scalar potential allowing unstable 
minima and construct  exact solutions of
 noncommutative false vacuum 
bubble\cite{strominger} 
with integer 
magnetic flux. We close our discussion by  mentioning  related
higher dimensional solutions and possible applications.

\noindent{\sl Vortices in Noncommutative Abelian-Higgs Theory:\ }
We shall begin with the noncommutative Abelian-Higgs model in 2+1 
dimensions described by\cite{jatcar,waida}
\begin{equation}
L=-{1\over g^2} \!\int\! d^2 x \Bigl( {1\over 4}F_{\mu\nu} * F^{\mu\nu}
+D_\mu\phi * (D^\mu\phi)^\dagger+{\lambda\over 2} 
(\phi *\phi^\dagger\!-\!v^2)^2\Bigr)
\label{lag}
\end{equation} 
where 
\begin{eqnarray} 
F_{\mu\nu}&=&{\partial_\mu} A_\nu\! -\!{\partial_\nu} A_\mu-i
(A_\mu\! *\! A_\nu\!-\!A_\nu\! *\! A_\mu)\nonumber\\
D_\mu \phi &=& \partial_\mu\phi-iA_\mu\!*\!\phi\,,
\end{eqnarray} 
and our metric convention is $\eta_{\mu\nu}={\rm diag}(-1,1,1)$.
Here the  
$*$-product is defined by
\begin{equation}
f(x)* g(x)\equiv \Bigl(e^{-i{\theta\over 2}\epsilon^{ij}\partial_i 
\partial'_j} f(x) g(x')\Bigr){\Big\vert}_{x=x'}\,,
\label{star}
\end{equation}
where we  take $\theta$ to be positive without loss of generality.
The theory can be equivalently presented by operators on the Hilbert space
defined by 
\begin{eqnarray}
[\hat{x}\,,\,\hat{y}]=-i\theta 
\end{eqnarray}
where the $*$-product
between functions becomes the ordinary product between the 
operators. 
For given 
function 
\begin{eqnarray}
f(x,y)=\int {d^2 k\over (2\pi)^2} \,
\tilde{f}(k) e^{i(k_x x+k_y y)}, 
\end{eqnarray}
the corresponding operator can be found
 by
the Weyl-ordered form of
\begin{eqnarray}
\hat{f}(\hat{x},\hat{y})=\int  {d^2 k\over (2\pi)^2} \,
\tilde{f}(k) e^{i(k_x \hat{x}+k_y \hat{y})}. 
\end{eqnarray} 
One may then easily show that 
$\int d^2x \,f$ is replaced by 
$2\pi\theta \,\tr \hat{f}$ and $\partial_i f$
corresponds to $-{i\over \theta}\epsilon_{ij}[ \hat{x}_j, \hat{f}]$. 
With the operator-valued fields,
the action can be written as 
\begin{equation}
L=-{2\pi\theta \over g^2} \tr [ \,{1\over 4}F_{\mu\nu} F^{\mu\nu}
+ D_\mu\phi(D^\mu\phi)^\dagger +{\lambda\over 2} (\phi\phi^\dagger-v^2)^2]
\label{lag1}
\end{equation} 
where hats are dropped for simplicity and the derivative notation
is understood as  $\partial_i f\equiv 
-{i\over \theta}\epsilon_{ij} [x_j, f]$.
The system is invariant under the gauge transformation,
\begin{eqnarray}
A_\mu'=U^\dagger A_\mu U +iU^\dagger\partial_\mu U\,, \ \ \
\phi'=U^\dagger\phi\,,
\label{gauge}
\end{eqnarray}
where the gauge group element $U$ satisfies 
\begin{eqnarray} 
U^\dagger U=U U^\dagger=I.
\end{eqnarray}
Using the translational invariance of the system,
the Hamiltonian can be constructed as\cite{jatcar}
\begin{eqnarray}
H={2\pi\theta \over g^2} \tr [ {1\over 2}(E^2+B^2)+D_t\phi (D_t\phi)^\dagger
+ D_i\phi (D_i\phi)^\dagger +{\lambda\over 2} (\phi\phi^\dagger-v^2)^2]\,,
\label{energy}
\end{eqnarray}
and the static equations of motion read
\begin{eqnarray}
&& \ \ D_iD_i\phi-\lambda  (\phi\phi^\dagger-v^2)\phi=0\,,\nonumber\\
&& \epsilon_{ij}D_jB=J_i\equiv i 
[\phi (D_i\phi)^\dagger -D_i\phi\, \phi^\dagger]\,, 
\label{eqofmotion}
\end{eqnarray}
where we use the gauge $A_0=0$ and set $\partial_t=0$.
Here the magnetic field 
transforms, under the gauge transformation, covariantly 
as $B'=U^\dagger B U$ and the covariant derivative on
an adjoint  field $H$ is  
understood as $D_\mu H=\partial_\mu H\!-\!i[A_\mu, H]$. 
Let us introduce the creation and annihilation operators by
$c^\dagger\equiv {1\over \sqrt{2\theta}}{(x\!+\!iy)}$ and by
$c\equiv {1\over \sqrt{2\theta}} 
{(x\!-\!iy)}$, which satisfy $[c,c^\dagger]=1$.  
To represent arbitrary operators in the Hilbert space we
shall use the occupation number basis by $G= \sum
g_{mn}|m\rangle\langle n|$
with the number operator $\hat{n}=c^\dagger c$. We will further denote
$A=A_x\!-\!iA_y$, $\,\partial_- G\equiv (\partial_x\!-\!i\partial_y) G=
{\sqrt{2}\over \sqrt{\theta}}[c,G]$
and  $\partial_+ G\equiv (\partial_x\!+\!i\partial_y) G=
-{\sqrt{2}\over \sqrt{\theta}}[c^\dagger,G]$.

We first present a unit flux static solution, where
the flux is defined by $\Phi=\theta \tr B$. (This corresponds to
$\Phi={1\over 2\pi}\int\! d^2x\, B$ when translated back to the language of 
ordinary functions.)  The unit flux solution is
\begin{eqnarray}
&& \phi=v\sum^\infty_{n=0}|n+1\rangle \langle n| 
\,,\nonumber\\
&&
 A={\sqrt{2}\over i\sqrt{\theta}}\left(c-{\sqrt{\hat{n}}\over 
\sqrt{\hat{n}+1}}c
\right)
={\sqrt{2}\over i\sqrt{\theta}}\sum^\infty_{n=0}
(\sqrt{n+1}-\sqrt{n})|n\rangle \langle n+1|\,.
\label{unitsolution}
\end{eqnarray}
It is simple to check that 
\begin{eqnarray}
B={1\over \theta}|0\rangle \langle 0|\,,
\end{eqnarray}
with which
$\Phi=1$. Furthermore $D_\pm\phi=0$ and $D_\pm B=0$, so the second 
equation of motion in (\ref{eqofmotion}) is satisfied. Using 
$D^2\phi= D_-D_+\phi - B\phi$,
one may show that $D^2\phi=0$ and $(\phi\phi^\dagger-v^2)\phi=0$, so the
first equation in (\ref{eqofmotion}) is also fulfilled. 

The flux number of solution corresponds to $\Phi
=1$. Thus the
solution describes a static localized vortex. 
Since $|n\rangle\langle n|$
is invariant under rotation, our solution is axially symmetric.
(Note here that $|n\rangle\langle n|$ is mapped to the function
$2(-1)^n L_n({2r^2\over \theta}) e^{- {r^2\over\theta}}$
where $L_n(z)$ is the n-th order Laguerre polynomial.)
By evaluating the 
Hamiltonian, one finds that the energy of the 
solution is 
\begin{eqnarray}
E_{\rm one}(v,\theta)={\pi\over g^2} 
\left({1\over \theta} +\lambda \theta v^4\right)\ge
{2\pi\over g^2}{\sqrt{\lambda}} v^2\,. 
\label{oneenergy}
\end{eqnarray}  
When $\lambda=1$, the theory allows so called Bogomol'nyi bound as discussed 
in Ref.~\cite{jatcar}. In fact it is straightforward to verify that the 
energy functional can be expressed as a complete squared form plus a
topological term by
\begin{eqnarray}
H={\pi\theta \over g^2} \tr [
(B+(\phi\phi^\dagger-1))^2+2(D_+\phi)(D_+\phi)^\dagger
+\epsilon_{ij}D_iJ_j+2 v^2 B] \ge {2\pi v^2\over g^2} \Phi\,,
\label{bound}
\end{eqnarray}
where we omitted the kinetic terms involving  $E_i$ 
and $D_t\phi$.
The saturation of the bound occurs once the Bogomol'nyi equations,
\begin{eqnarray}
D_+\phi=0,\ \ \  B=1-\phi\phi^\dagger\,, 
\end{eqnarray}
are satisfied. 
When $\lambda=1$,
the bound in (\ref{oneenergy}) agrees with the Bogomol'nyi bound
that is an absolute energy  bound for one vortex solution. Hence
when $v^2={1\over \theta}$ and $\lambda=1$, the solution should be 
a BPS solution. Indeed for the specific value of $v$, one can check
that the solution satisfies the BPS equations. 
This BPS solution is clearly  stable because they
saturate the energy bound set by the topological quantity.
For $\lambda\neq 1$ or $v^2\neq {1\over \theta}$, 
it is not clear at this point 
whether  these solutions are
classically stable or not. 
For generic $\lambda$, 
the saturation of the bound 
in (\ref{oneenergy}) can be achieved when  
$v^2=1/(\sqrt{\lambda}\theta)$, whose physical 
implications are again not clear.

There are some physically  equivalent configurations
connected by gauge transformations. 
For example, $B={1\over \theta}
|k\rangle \langle k|$ solution with $k>0$ can be obtained by the 
gauge transformation with 
\begin{eqnarray}
U=1\!-\!
|k\rangle \langle k|\!-\!|0\rangle \langle 0|\!+\!|k\rangle \langle 0|
\!+\!
|0\rangle \langle k|.  
\end{eqnarray}
Also rather simple extension of
above solution comes from the translational invariance of the system.
Our solution describes a vortex located at the origin
and  
the explicit form  of vortex  at $(a_x,a_y)$,
can be constructed with help of the translation generators $T_i=
-{1\over\theta}\epsilon_{ij}[x_j,\ \!]$. For the above solution or 
solutions below, they will be just given as
$A'=e^{-ia_iT_i}A$ and $\phi'=e^{-ia_iT_i}\phi$.

We now turn to the case of $\Phi =m\,(m\ge 1\,, m\in Z)$ vortex 
solutions. The solutions read
\begin{eqnarray}
&&\phi=v\sum^\infty_{n=0}|n+m\rangle \langle n| 
\,,\nonumber\\
&& A={\sqrt{2}\over i\sqrt{\theta}}\Bigl(c\!-\!K_m
{\sqrt{\hat{n}\!\!-\!\!m\!\!+\!\!1}\over \sqrt{\hat{n}\!+\!1}}K_m\, c
\Bigr)
=\!{\sqrt{2}\over i\sqrt{\theta}}\sum^\infty_{n\!=\!0}
\sqrt{n\!\!+\!\!1}\left(|n\rangle \langle n\!\!+\!\!1|
\!-\!
|n\!\!+\!\!m\rangle \langle n\!\!+\!\!m\!\!+\!\!1|
\right)\,,
\label{multisolution}
\end{eqnarray}
where the operator $K_m$ denotes the projection operator
$1-\sum_{n=0}^{m-1}|n\rangle\langle n|$. Similar to the case of one
 vortex, the static  equations of motion can be checked explicitly.
Especially, they again satisfy $D_\pm \phi =0$ and the magnetic field
 $B$ is given by 
\begin{eqnarray}
B=\sum_{n=0}^{m-1}|n\rangle\langle n|.
\end{eqnarray} 
Thus the 
solution carries the vortex number $\Phi=m$. The energy for the 
solution is evaluated as  
\begin{eqnarray}
E_m= m E_{\rm one}. 
\end{eqnarray}
The saturation of the
Bogomol'nyi bound occurs at $v^2={1\over \theta}$ with $\lambda=1$
and, for these values of parameters, one may check that
the solutions satisfy the BPS equations. 

Interestingly, the above gauge field $A$ 
is also static solution of the pure gauge 
theory\cite{polychronakos}, where the
static equation reads $D_i B=0$.
We note that the magnetic field of the solution is
well localized with a size of order $\sqrt{\theta}$. This kind of 
localized  pure magnetic solutions cannot 
be static in the ordinary pure U(1) 
theory. This is because nothing prevents
the magnetic field from spreading out, which will result in a 
configuration of the lower energy in the ordinary U(1) 
Maxwell theory. 

A few comments are in order. First of all,
these multi-vortex solutions are  static. For the case when 
the Bogomol'nyi bound is saturated, the existence of static 
vortex solution has clear physical reasons. In the commutative 
case, there is of course no long range interactions 
because the Higgs and photons become massive by the Higgs mechanism. 
However, the vortices  even may overlap completely and, hence,
the cancellation of short-ranged interaction is necessary in order 
to
have  static configurations.
When the BPS equations are satisfied,  the  
forces between the vortices cancel out indeed.  
Namely, when we have  vortices, there is in general
current circulating around the vorticities. Because there
is also local magnetic field, these combine to give
 local Lorenz force 
density. Also there is force contribution from Higgs
gradient and the Higgs potential. These contributions
 should be canceled 
whenever we 
have a static configuration of vortices. 
The BPS equations may be regarded as such conditions that guarantee
the cancellation 
of the local force density.
In case of the static noncommutative multi-vortex, 
the current density $J_i$ is zero and, also,
the contributions from the Higgs gradient and potential
vanish since $(D_i\phi) \phi^\dagger=0$
and $(\phi\phi^\dagger\!-\!1)\phi=0$.
Hence the existence of the static multi-vortex solutions indeed
relies upon
vanishing of local force densities.
Secondly, one may alternatively consider the theory
with $V={ \lambda\over 2} \tr 
(\phi^\dagger\phi- v^2)^2$, witch  is still gauge invariant.
It is then simple to show that
 the solution above solves again  this new theory.
However the energy
will become different. This is due to the fact
that $\tr (\phi^\dagger\phi- v^2)^2\neq 
\tr (\phi\phi^\dagger- v^2)^2$ in general.
Thirdly we can check explicitly that the topological number
$\Phi (=m)$ of the multi-vortex 
agrees with the index\cite{jatcar,furuuchi,witten} 
\begin{eqnarray}
I\equiv {\rm dim}[{\rm ker}\, \phi^\dagger]\!-\!
{\rm dim}[{\rm ker}\, \phi]
\end{eqnarray}
on our solutions.

\noindent{\sl Noncommutative False Vacuum Bubble with Magnetic Flux:\ }
Recently, the localized vacuum bubble solitons are found
in noncommutative field theories\cite{strominger}. They
are localized false vacuum bubble of
the relevant theory, where the noncommutativity
of coordinates prevents its collapse to a zero size
as  dictated the uncertainty relation 
$\Delta x\Delta y\ge \theta$. Thus the bubble has a size of 
order $\theta$ and, outside of false vacuum  region of core,
the field configuration  approaches to its stable 
minimum.  Until now, no closed form solutions are found for finite 
$\theta$\cite{finite}.

We shall present similar false vacuum bubble solutions in the 
Abelian-Higgs model with $m$ magnetic flux. 
To have false vacuum bubbles, we consider 
the system in (\ref{lag1}) but with more
general potentials of
 the form ${2\pi\theta\over g^2}V(\xi)$
with $\xi\equiv \phi\phi^\dagger$ 
that is consistent with the
gauge symmetry.
We further assume that the true minima of 
potential are located at $\xi
=0$ with $V(0)=0$
and there is at least one other local minimum.
The static equations of motions are  given by 
\begin{eqnarray}
 D_iD_i\phi- V'(\phi\phi^\dagger)\phi=0\,,
\label{falseeq}
\end{eqnarray} 
with $\epsilon_{ij} D_jB=J_i$ that is the same as before.
The static solutions with $m \,(\ge 1)$ magnetic flux are given by
\begin{eqnarray}
\phi&=&\phi_0 |0\rangle \langle 0| 
\,,\nonumber\\
 A&=& -i{\sqrt{2}\over \sqrt{\theta}}\Bigl(c- e^{i\nu}K_m
{\sqrt{\hat{n}\!\!-\!\!m\!\!+\!\!1}\over \sqrt{\hat{n}\!+\!1}}K_m\, c
\Bigr)\,,
\label{strosolution}
\end{eqnarray}    
where $\nu$ is real constant and $\phi_0$ is determined by an 
algebraic
equation by 
\begin{eqnarray}
1+ \theta V'(\xi_0)=0\,.
\label{minimum}
\end{eqnarray} 
with $\xi_0=|\phi_0|^2$.   
To show that they are indeed satisfying the equations 
of motion, we note that  $B=\sum_{n=0}^{m-1}|n\rangle\langle n|$, 
$D_+\phi=0$
and $D_-\phi=-{\sqrt{2}\over \sqrt{\theta}}\phi_0 |
0\rangle \langle 1|$. Because $(D_-\phi) \phi^\dagger=0$,
and $J_i=D_iB=0$. The equation of motion in (\ref{falseeq})
leads to the  
condition of (\ref{minimum}).
The flux number is $\Phi=m$ and the energy is given 
by
\begin{eqnarray}
E={2\pi\over g^2} \left({m\over 2\theta}
+\xi_0
+ \theta V(\xi_0
)
\right)\,.
\label{stroenergy}
\end{eqnarray}     
 When $\theta$ is large enough,
the values of $
\xi_0$ satisfying the condition
 is given by the extremum of 
the potential, which is expected from the analysis
of Ref.~\cite{strominger} for the scalar noncommutative 
solitons. Also the leading term of energy
in the large $\theta$ limit is again consistent with
the analysis given in Ref.~\cite{strominger}.

Although one may consider
more general potentials easily, we illustrate further 
properties of the solutions with 
\begin{eqnarray}
V(\xi)=\beta [
2\xi^3-3(a+b)\xi^2+ 6ab\xi]\,, 
\end{eqnarray}
where we take  $b>a>0$. 
Noting $V'(\xi)=6\beta (\xi-a)
(\xi-b)$. The condition (\ref{minimum}) becomes 
now $1+6\theta \beta (\xi-a)(\xi-b)=0$. Hence
when $\theta\rightarrow \infty$, the solutions are
$|\phi_0|^2=a, b$. Here $a$ corresponds to 
the classically unstable soliton, while $b$ gives 
a classically stable one. For general $\theta$,
the solutions read
\begin{eqnarray}
2\xi^\pm_0={a\!+\!b \pm \!\sqrt{(b\!-\!a)^2\!-\!{2\over 3\theta\beta}}}
\end{eqnarray}
for $(b\!-\!a)^2\!-\!{2\over 3\theta\beta}\ge 0$. Otherwise, there are 
no solutions satisfying the condition.
The branch $\xi^-_0$ is classically 
unstable because it maximizes the energy 
in (\ref{stroenergy}).


Finally, one may find more general solutions as 
follows. Let us consider the Higgs and gauge fields of the form,
\begin{eqnarray}
\phi&=&\sum^{m-1}_{k=0}\phi_k |k\rangle \langle k| 
\,,\nonumber\\
 A&=& -i{\sqrt{2}\over \sqrt{\theta}}\Bigl(c- e^{i\nu}K_m
{\sqrt{\hat{n}\!\!-\!\!m\!\!+\!\!1}\over \sqrt{\hat{n}\!+\!1}}K_m\, c
\Bigr)\,.
\label{stroaa}
\end{eqnarray}    
The static equations of motion (\ref{falseeq}) are then reduced to
\begin{eqnarray}
&& 
J_+ ={\sqrt{2}\over i\sqrt{\theta}}
\sum^{m\!-\!2}_{k=0}\big[\phi_{k\!+\! 1}\phi^*_{k} |k\!+\!1\rangle \langle k| -
\phi_k\phi^*_{k\!+\!1} |k\rangle \langle k\!+\!1|\big]=0 
\,,\nonumber\\
&&\ \ \ \ \ \   
\big[2k+1+ \theta V'(\xi_k)\big]\,\xi_k=0 \,,
\label{strobb}
\end{eqnarray}  
where $J_+\equiv J_1+iJ_2$ and $\xi_k=|\phi_k|^2$.
The first equation implies that $\phi_k\phi^*_{k+1}=0$ for all $k$.
Namely, for any given nonvanishing  $\phi_k$, the neighboring 
elements, $\phi_{k\pm1}$ ought to be zero. The second 
equation is again purely 
algebraic and can be solved once the explicit form of the 
potential is given.

In this note,  we construct exact multi-vortex solutions of 
the noncommutative Abelian Higgs theory.
 Further considering the Abelian-Higgs 
theory with more complicated scalar potential, we find static
false vacuum bubble solutions with $m$ magnetic flux.
The mechanism behind the existence of
 such a static multi-vortex solution
is not fully identified. Hence the way we get the solutions
was not quite systematic. In this respect, it is 
not clear whether
we have obtained full category of
such solutions. This requires further study. 

Not to mention, the low energy moduli dynamics of 
our multi-solitons
will be particularly of interest.
Similar multi-vortex solutions 
are also expected in
the noncommutative version of the nonrelativistic scalar 
theory\cite{jackiw} 
 or the relativistic gauge 
theory\cite{yoonbai} coupled to  Chern-Simons gauge theory
and a work is in progress\cite{bak}.
It is also quite obvious that,
$D+1$ dimensional Abelian-Higgs type theory,
 trivial embedding of our solutions will give extended objects 
of dimension $D-2$. Our solutions
 would  also be  relevant to understanding
the tachyon condensation in string theory\cite{witten,harvey}.  
 
\vskip .3cm
\noindent{\bf Note added}: The detailed stability analysis 
of the pure flux solutions in the U(1) Maxwell theory
has recently appeared in \cite{aganagic}. In this paper,
it is shown that there are tachyonic modes in 
their fluctuation spectra.

\vskip .3cm
\noindent{\bf Acknowledgment}
This work 
is supported
in part by KOSEF 1998 Interdisciplinary Research Grant
98-07-02-07-01-5. We like to thank the enlightening discussions with
K. Lee and J. H. Yee.
We also would like to thank the hospitality of YVRC at Yonsei
University where part of this work is done.

\end{document}